\documentclass[twocolumn]{aastex63}
\usepackage{amsmath}
\usepackage{CJK}

\newcommand{\uatnum}[1]{\href{http://vocabs.ands.org.au/repository/api/lda/aas/the-unified-astronomy-thesaurus/current/resource.html?uri=http://astrothesaurus.org/uat/#1}{#1}}
\newcommand{\northwestern}{Center for Interdisciplinary Exploration and Research in Astrophysics (CIERA), Northwestern University, 1800 Sherman, Evanston, IL,
9 60201, USA}
\newcommand{\caltech}{Department of Astronomy, California Institute of Technology, Pasadena, CA 91125, USA}
\newcommand{\ucsc}{Department of Astronomy \& Astrophysics, University of California, Santa Cruz, CA95064, USA}
\newcommand{\keck}{W. M. Keck Observatory, 65-1120 Mamalahoa Hwy, Kamuela, HI, USA}
\newcommand{\ucla}{Department of Physics \& Astronomy, 430 Portola Plaza, University of California, Los Angeles, CA 90095, USA}
\newcommand{\jpl}{Jet Propulsion Laboratory, California Institute of Technology, 4800 Oak Grove Dr.,Pasadena, CA 91109, USA}
\newcommand{\ucsd}{Center for Astrophysics and Space Sciences, University of California, San Diego, La Jolla, CA 92093, USA}
\newcommand{\arizona}{James C. Wyant College of Optical Sciences, University of Arizona,
Meinel Building 1630 E. University Blvd., Tucson, AZ. 85721}

\submitjournal{AAS Journals}

\shorttitle{$\kappa$ And b with KPIC}
\shortauthors{Morris et al.}

\begin{document}
\begin{CJK*}{UTF8}{gbsn}

\title{$\kappa$ And b is a fast rotator from KPIC High Resolution Spectroscopy}
\correspondingauthor{Evan Morris}
\email{ecmorris@ucsc.edu}

\author[0000-0003-3165-0922]{Evan C. Morris}
\affiliation{\ucsc}

\author[0000-0003-0774-6502]{Jason J. Wang (王劲飞)}
\affiliation{\northwestern}

\author[0000-0002-5370-7494]{Chih-Chun Hsu}
\affiliation{\northwestern}

\author[0000-0003-2233-4821]{Jean-Baptiste Ruffio}
\affiliation{\ucsd}

\author[0000-0002-6618-1137]{Jerry W. Xuan}
\affiliation{\caltech}

\author[0000-0001-8953-1008]{Jacques-Robert Delorme}
\affiliation{\keck}
\affiliation{\caltech}

\author[0000-0003-1150-7889]{Callie Hood}
\affiliation{\ucsc}

\author[0000-0002-6076-5967]{Marta L. Bryan}
\affiliation{David A. Dunlap Institute Department of Astronomy \& Astrophysics, University of Toronto, 50 St. George Street, Toronto, ON M5S 3H4, Canada}

\author[0000-0002-0618-5128]{Emily C. Martin}
\affiliation{\ucsc}

\author{Jacklyn Pezzato}
\affiliation{\caltech}

\author[0000-0002-8895-4735]{Dimitri Mawet}
\affiliation{\caltech}
\affiliation{\jpl}

\author[0000-0001-6098-3924]{Andrew Skemer}
\affiliation{\ucsc}

\author[0000-0002-6525-7013]{Ashley Baker}
\affiliation{\caltech}

\author{Randall Bartos}
\affiliation{\jpl}

\author[0000-0003-4737-5486]{Benjamin Calvin}
\affiliation{\ucla}

\author{Sylvain Cetre}
\affiliation{\keck}

\author{Greg Doppmann}
\affiliation{\keck}

\author[0000-0002-1583-2040]{Daniel Echeverri}
\affiliation{\caltech}

\author[0000-0002-1392-0768]{Luke Finnerty}
\affiliation{\ucla}

\author[0000-0002-0176-8973]{Michael P. Fitzgerald}
\affiliation{\ucla}

\author[0000-0001-5213-6207]{Nemanja Jovanovic}
\affiliation{\caltech}

\author[0000-0002-4934-3042]{Joshua Liberman}
\affiliation{\arizona}

\author[0000-0002-2019-4995]{Ronald Lopez}
\affiliation{\ucla}

\author[0000-0003-1399-3593]{Ben Sappey}
\affiliation{\ucsd}

\author{Tobias Schofield}
\affiliation{\caltech}

\author[0000-0001-5299-6899]{J. Kent Wallace}
\affiliation{\jpl}

\author[0000-0002-4361-8885]{Ji Wang (王吉)}
\affiliation{Department of Astronomy, The Ohio State University, 100 W 18th Ave, Columbus, OH 43210, USA}

\begin{abstract}
We used the Keck Planet Imager and Characterizer (KPIC) to obtain high-resolution (R$\sim$35,000) K-band spectra of $\kappa$ Andromedae b, a planetary-mass companion orbiting the B9V star, $\kappa$ Andromedae A. We characterized its spin, radial velocity, and bulk atmospheric parameters through use of a forward modeling framework to jointly fit planetary spectra and residual starlight speckles, obtaining likelihood-based posterior probabilities. We also detected H$_{2}$O and CO in its atmosphere via cross correlation. We measured a $v\sin(i)$ value for $\kappa$ And b of $38.42\pm{0.05}$ km/s, allowing us to extend our understanding of the population of close in bound companions at higher rotation rates. This rotation rate is one of the highest spins relative to breakup velocity measured to date, at close to $50\%$ of breakup velocity. 
We identify a radial velocity $-17.35_{-0.09}^{+0.05}$ km/s, which we use with existing astrometry and RV measurements to update the orbital fit. We also measure an effective temperature of $1700{\pm100}$~K and a $\log(g)$ of $4.7{\pm0.5}$~cgs dex.
\end{abstract}


\section{Introduction}\label{sec:intro}
New capabilities in high resolution ($R > 10,000$) spectroscopy open avenues for further exploration of the orbital and atmospheric parameters of directly imaged planetary-mass companions. In particular, this resolving power allows us to look at the shapes and shifts of individual absorption lines in the planetary and substellar atmospheres to understand properties such as rotation and radial velocity~\citep[e.g.][]{Snellen2014,Ruffio2019,Bryan2020,Wang2021}. Further, forward modeling and cross correlation between data and models gives us insight into temperatures and compositions~\citep[e.g.][]{Konopacky2013,Brogi2019,Wang2022,Xuan2022,Landman2023}.

Previously, these investigations have been primarily limited to brighter companions located at greater distances from their stars, as the instrumentation used for this purpose utilized slit spectroscopy. In these cases, at small angular separation, stellar light would contaminate the data to the point of overwhelming the planetary signal. 
This problem can be mitigated using Keck Planet Imager and Characterizer~\citep[KPIC;][]{Mawet2017,Jovanovic2019,Delorme2021}, which is designed to combine single-mode fiber, high resolution spectroscopy with high contrast imaging techniques, the combination of which has been shown to significantly decrease the stellar glare present in the data and improve post-processing possibilities~\citep{Wang2021}. This combination of technologies allows for observation of fainter planets at smaller angular separation from their stars.

To date, KPIC has been used to expand NIRSPEC's high-resolution spectroscopy to companions that are smaller separation (typically at or within $\sim$1"), particularly looking at the spin parameter, radial velocity, and atmospheric information, including molecular detections, abundances and effective temperature, for the HR 8799 planets~\citep{Wang2021,Wang2023}, HR 7672 B~\citep{Wang2022}, HD 4747 B~\citep{Xuan2022}, HIP~55507~B~\citep{Xuan2024}, and HD 33632 Ab~\citep{HsuNEW}.

Understanding planetary spin allows us insight into the system's formation history and evolution. Planetary mass objects spin up and conserve angular momentum after disk dispersal~\citep{Bryan2020}. Recent work shows that companions spin at $10\%$ of breakup velocity generally, and this can be explained by magnetic breaking in the early phases, in the  release of angular momentum caused by interaction with the magnetized circumplanetary disk~\citep{Batygin2018, Bryan2020}. There is related tentative evidence of an anti-correlation between spin and companion mass, pending access to a larger sample size of low-mass companion spin measurements~\citep{Batygin2018, Wang2021, HsuNEW}.

We focus our analysis on $\kappa$ Andromedae b \citep{Carson2013}. This planetary mass companion orbits a B9V star, understood to be a potential member of the Columba association, with an age of $\sim$30 Myr \citep{Zuckerman2011}, leading to a mass estimate of $\sim$12.8 $M_\textrm{Jup}$ in \citet{Carson2013}. However, \citet{Hinkley2013} found a much older isochronal age for this object, $220 \pm 100$ Myr, corresponding to a mass of $50_{-13}^{+16}$ $M_\textrm{Jup}$. Following this, \citet{Jones2016} used CHARA interferometry to constrain properties of $\kappa$ And A to compare to evolutionary models, finding that the models favored a younger age of $47_{-40}^{+27}$ Myr, which is in agreement with the most recent Columba Association age estimate of $42_{-4}^{+6}$ Myr from \citet{Bell2015}.

Other studies of $\kappa$ And b indicate that the object is low gravity (with a $\log(g)$ of 4--4.5 cgs dex), in agreement with models suggesting a lower age \citep{Currie2018,Uyama2020}. \citet{Uyama2020} and \citet{Currie2018} find a temperature range of 1700--2000 K, \citet{Wilcomb2020} finds a range of 1950--2150 K. This appears to be dependent on selected model grid, and the range in \citet{Uyama2020} is found when comparing 13 grids, including the one we chose for this paper. \citet{Wilcomb2020} derives a C/O ratio of $0.70_{-.24}^{+0.09}$ for $\kappa$ And b, which, when paired with their subsolar metallicity measurement, suggests consistency with the host star and rapid formation via gravitational instability. \citet{Currie2018} estimates a mass of $13_{-2}^{+13}$ $M_\textrm{Jup}$, which we use in further analysis.\footnote{The literature defines $\kappa$ And b variably, sometimes as a planetary-mass companion, bound companion, super-Jupiter, or low-mass brown dwarf~\citep{Carson2013,Uyama2020,Wilcomb2020,Hinkley2013,Stone2020}. It is likely near or just below the limit for deuterium burning~\citep{Carson2013,Hinkley2013}, but its orbital inclination may allow for disk formation by disk instability~\citep{Currie2018}. We have chosen to refer to it as a planetary-mass companion or a planet.}

Section \ref{sec:obs} describes the observation sequence of $\kappa$ And b performed with KPIC. Section \ref{sec:dr} details the data reduction process developed for this dataset, and used broadly for KPIC observations. In Section \ref{sec:fit}, we describe our forward modeling framework for fitting atmospheric models to the data, as well as discuss our cross correlation molecular detections of H$_2$O and CO in the planet's atmosphere. Section \ref{sec:disc} contextualizes our measurements, including the first spin for this object, a new orbital fit based on additional radial velocity data, and analysis of bulk atmospheric parameters. We finish in Section \ref{sec:con} by summarizing our work, and discussing future avenues of improvement.

\section{Observations}\label{sec:obs}

\begin{figure*}
    \centering
    \includegraphics[width=0.8\textwidth]{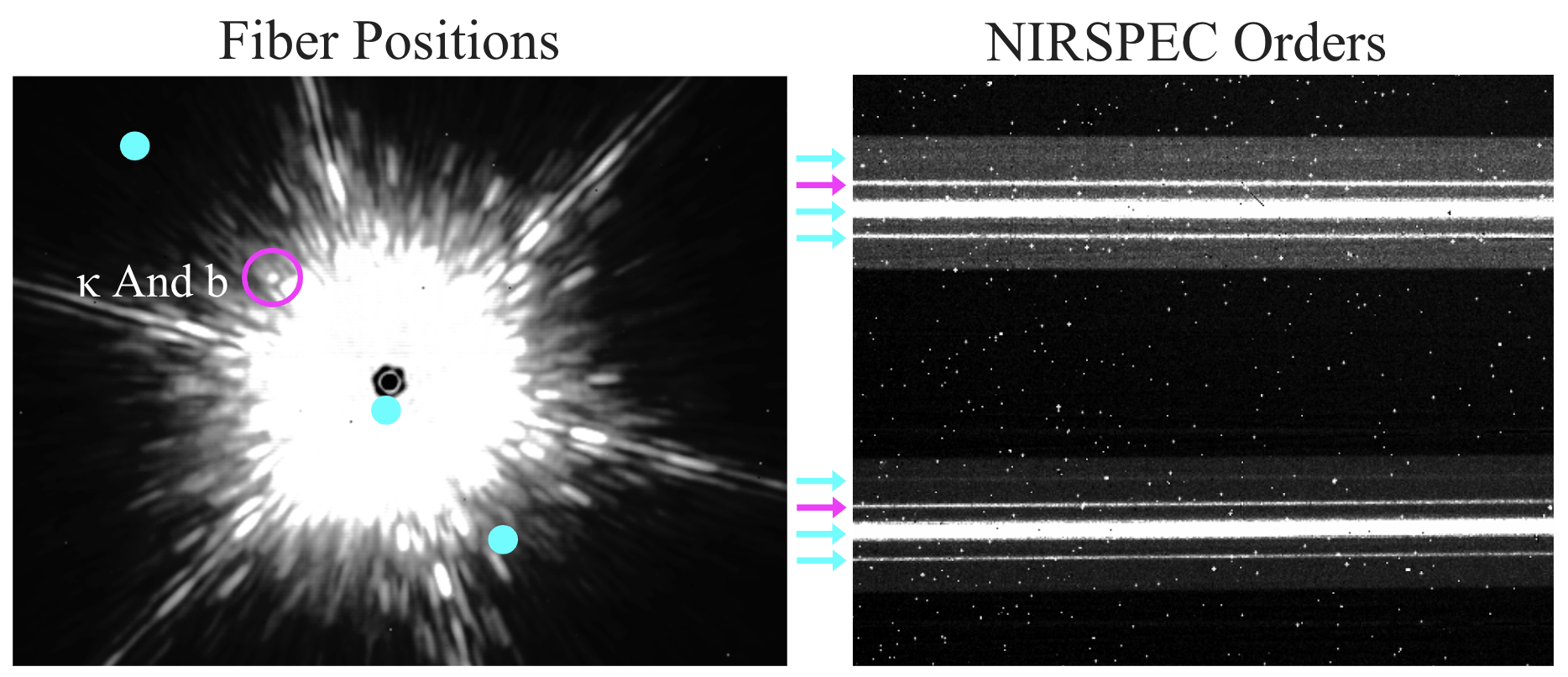}
    \caption{
Visualization of the radial location of fibers, shown on sky and on the NIRSPEC detector when observing with the KPIC fiber injection unit. One fiber is placed on the companion, one fiber is placed near the star, and the other fibers capture varying amount of stellar light. This NIRC2 image was not taken during the observing sequence, but serves as an example of the dynamic range of the stellar halo, though a coronagraph is not used during KPIC observations, so there is a slight difference. The right-hand image is a small portion of two echelle orders on the NIRSPEC detector, showing a portion of the fiber traces in those orders.}
    \label{fig:onsky}
\end{figure*}

\subsection{Instrument Description}
The Keck Planet Imager and Characterizer (KPIC) is an instrument optimized for obtaining high-resolution spectra of directly imaged exoplanets. KPIC includes upgrades to the Keck II adaptive optics system, paired with a fiber injection unit (FIU) to Keck’s existing high-resolution near-infrared spectrograph, NIRSPEC (R$\sim$35,000)~\citep{McLean1998,Martin2018}. Light is coupled into an array of single mode fibers, which are fed into the NIRSPEC slit by the fiber extraction unit (FEU). These fibers allow KPIC to spatially separate light from a planet from that of its host star. See~\citet{Delorme2021} for an in-depth instrument summary.

\subsection{Instrument Setup}
When taking data, we direct light through KPIC into NIRSPEC. We use a custom pupil stop in NIRSPEC, as the FEU creates a different optical beam than that of NIRSPEC when it is used directly behind the telescope or AO system. We use NIRSPEC's “Thin” filter, clear PK50 glass that is used to suppress red-leaks in JHK filters by blocking wavelengths longer than $\sim2.5$ $\mu$m. Because the FEU pupil is different, we offset the PK-50 from it's nominal position to avoid vignetting from a central obscuration on the PK-50 substrate, though since these observations, new K-band filter with better K-band transmission and no central obscuration have been installed in NIRSPEC. The NIRSPEC echelle grating was set to $63.0^\circ$ and the cross-disperser was set to $35.76^\circ$, resulting in data containing nine spectra orders, NIRSPEC orders 31-39, ranging from approximately 1.94 to 2.49 $\mu$m. The disperser positions were held fixed for all of the data and calibrations described in this paper.

\subsection{\texorpdfstring{$\kappa$}{k} And b observations}
On UT July 3 2020, we observed $\kappa$ And b in K band using KPIC.

We began our observing sequence by designating a primary science fiber, based on which fiber had the best end-to-end throughput in daytime testing, as well as identifying which fiber was closest to the host star during our science observations, designating that fiber as our secondary. Using the adaptive optics field rotator (K-mirror), we rotated the field of view relative to our fixed fiber bundle such that this secondary fiber captured simultaneous stellar spectra while we observed the companion on the primary fiber, but was not directly aligned closely enough to saturate on the host star during the longer exposure. These simultaneous stellar spectra were used in data exploration, but on-axis observations of the star were used for our final analysis.

We first placed the host star, $\kappa$ And A, on the primary and secondary science fibers, taking three 30~s exposure in each position, for telluric calibration purposes. Using the FIU, we offset the star such that $\kappa$ And b is placed on the primary science fiber, as shown in Figure \ref{fig:onsky}. We then took 600~s exposures with NIRSPEC, capturing light from the companion, as well as scattered starlight, on the primary fiber, and significant starlight on the secondary fiber. Our remaining on-sky fibers also captured some starlight, as well as background. After our first hour of observations on the companion, we returned to the host star and repeated our calibration exposures before returning to the companion.

Our total integration time on $\kappa$ And b was 100 minutes with ten 600 s exposures, over a range in airmass from 1.1 to 1.25.

\subsection{Calibration Data}
In addition to calibrating using the spectra of the host star obtained during the observing sequence, we derive our wavelength solution by observing an M-giant star. We observed M giant HIP 81497 (spectral type M2.5III), and telluric standard star ups Her (spectral type B9III), which we observed twice during the night to check for wavelength solution shifts due to an M-4.6 earthquake near Maunakea during the night. Because all data on $\kappa$ And b was taken afterwards, the calibration data after the quake was used for all further analysis, though any shift was minimal. For these calibrations, we took five 1.5~s exposures of HIP 81497 and three 30~s exposures of ups Her on-axis in each of the four science fibers on slit that night. The dominant thermal noise in our data is from NIRSPEC, so after the end of the night, we took a series of thermal background frames at each exposure time we used (600 s for science images, 1.5 s and 30 s for calibration images), for use in background subtraction and in identifying bad pixels. These capture the thermal background of KPIC when no light is being injected into any of the fibers, in order to measure thermal emission from  KPIC~\citep{Wang2021}.

\section{Data Reduction}\label{sec:dr}

\begin{figure*}
    \centering
    \includegraphics[width=0.9\textwidth]{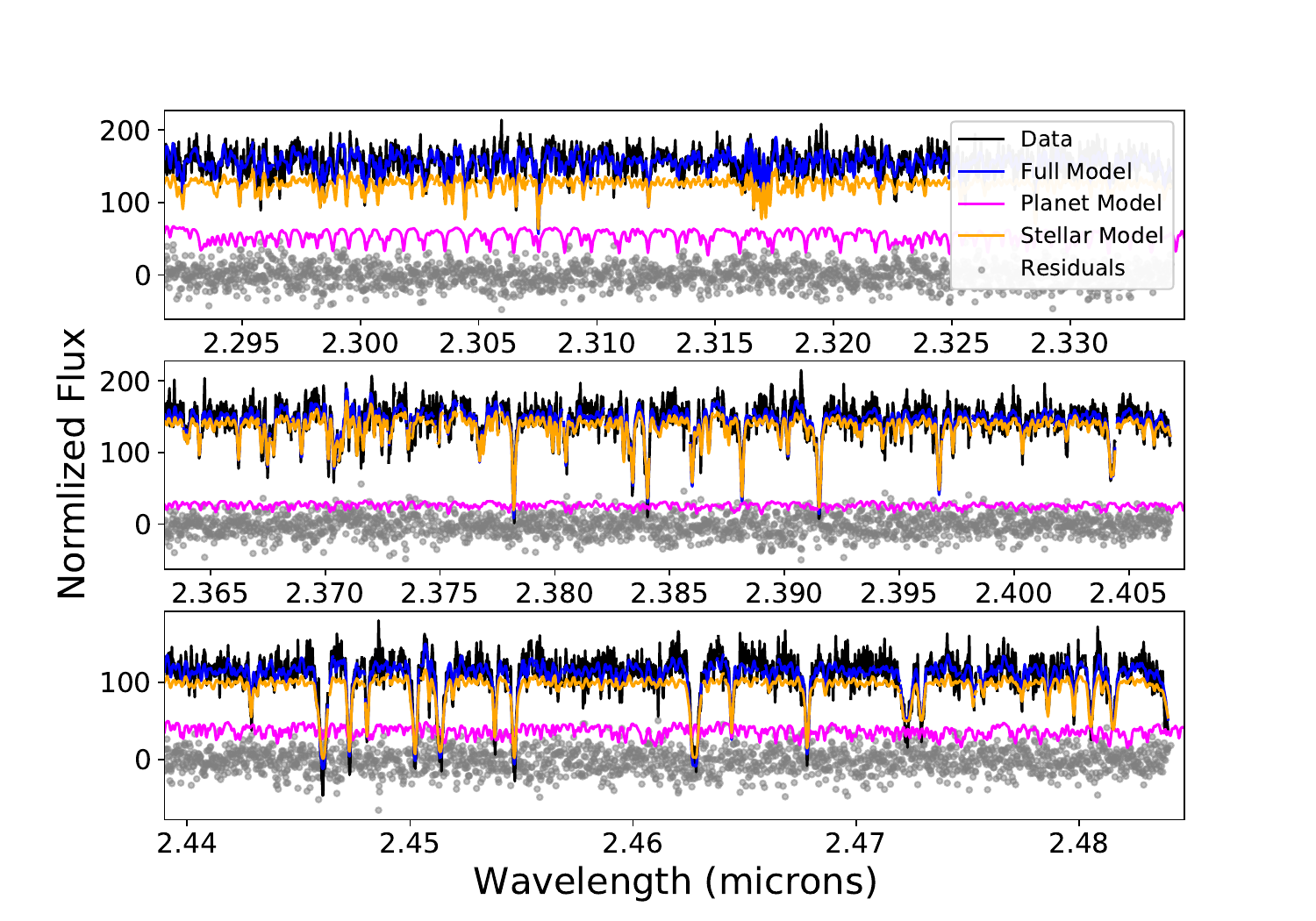}
    \caption{Extracted spectra in NIRSPEC orders 31-33 used in the fit compared to the best fitting forward model for that order, with contributions from companion and speckle flux also shown independently.}
    \label{fig:bestfit}
\end{figure*}

\subsection{Raw Data Reduction and Extraction}
We reduce our data and extract spectra using a custom-built pipeline for KPIC data\footnote{\url{https://github.com/kpicteam/kpic_pipeline}}. Our data is produced by NIRSPEC, so it shares size and format with other NIRSPEC data taken since its most recent upgrade (2048 by 2048 pixels, NIRSPEC’s order layout)~\citep{Martin2018}, but KPIC’s fiber-fed configuration allows and requires us to modify the reduction process accordingly to take advantage of the stable traces and line spread function of the single mode fiber.

Using the thermal background frames with exposure times coordinated to that of our science and calibration data, we create combined background frames using the mean of each set. We find the noise in each pixel by taking the standard deviation of values across all frames, and use these to locate bad pixels on the NIRSPEC detector.

We subtract the appropriate background frames from our science and calibration data, and mask identified bad pixels. These do not perfectly subtract off the background in our data, as the FEU is at a different temperature during the day, when our background frames are taken. This residual background is modeled in the extraction step. 

Instead of rectifying and extracting each order, we use data on the telluric standard to locate each fiber on the detector, allowing us to extract 1D spectra directly. The single mode fibers we use produce a point spread function (PSF) on the detector that is approximately a 2D Gaussian. By measuring the position and standard deviation of the PSF at each vertical column of the detector, we are able to identify each trace, fitting a spline to smooth these measured positions.

Before extracting, we account for the residual background leftover after an imperfect background subtraction, likely due to internal temperature differences within the instrument between daytime backgrounds and nighttime observations. We offset from the measured fiber trace, using pixels at least five pixels away from the center of a fiber, to estimate the remaining background in each column. We subtract the median of these pixels from each pixel in the column.

We use the trace calibration derived from the telluric standard to extract the spectra of in every column of each fiber. We use optimal extraction~\citep{Horne1986} to measure the flux using a 1D Gaussian profile as the PSF, with positions and standard deviations determined by the trace measurements taken on the standard star. We fit for the amplitude of the Gaussian and convert amplitude to total flux, weighing each pixel by the noise computed from the background frames and photon noise from the data itself. The flux in each column of the fiber is the total integrated flux of this 1D Gaussian, using the uncertainty of the optimal extraction as the uncertainty in the flux measurement~\citep{Horne1986}. We repeat this measurement across the full slit and for every order, resulting in 9 orders of extracted 1D spectra with 2048 channels per order. We also measure the instrumental spectral response of each science fiber.

\subsection{Wavelength Calibration}
To derive a wavelength solution, we use data on each fiber from a bright M-giant star, HIP 81497, and compare the observed spectra to models. We compute the expected RV shift of the star using barycentric correction and the systemic RV. We use a PHOENIX-ACES-AGSS-COND-2011 stellar model spectrum with a temperature of 3600 K~\citep{Husser2013}, an ATRAN\footnote{\url{https://atran.arc.nasa.gov/cgi-bin/atran/atran.cgi}}~\citep{Lord1992} telluric transmission model of the atmosphere, and the spectral response of instrument to build a model of the starlight, adjusted for continuum and background. For each order, we fit for wavelength using a cubic spline function, first performing a grid search with three nodes for an initial fit, and then running a non-linear simplex optimizer (Nelder-Mead) to identify a best fit.

\section{Fitting High-Resolution Spectra}\label{sec:fit}

\subsection{Forward Modeling \texorpdfstring{$\kappa$}{k} And b Spectra}
When fitting our spectra of $\kappa$ And b, we create a forward model of our data using the following method, defined in~\citet{Wang2021}. 

We directly compare our extracted spectra to a forward model containing all expected companion, stellar, telluric, and background contributions~\citet{Ruffio2019}. We also compare results to a model in which we include no signature of a companion as a baseline.

In order to create this model, we need to understand what light ends up in our fiber and on our detector, with contributions from the star, planet, atmosphere, telescope, and instrument. 

Data from a fiber placed on a planet can described as follows \citep{Wang2021}: 
\begin{equation}
    D_p(\lambda) = \alpha_p(\lambda) T(\lambda)P(\lambda) + \alpha_s(\lambda) T(\lambda)S(\lambda) + n(\lambda).
\end{equation}

In this equation, $D_p$ represents light extracted from the planet fiber. $T$ is optical system transmission, including components from the atmosphere, telescope, and instrument. $P$ is the flux from the planet and $S$ is the flux from the star, while $\alpha_p$ is the coupling efficiency of planet light and $\alpha_s$ is the coupling efficiency of stellar speckles into the fiber. $n$ is the noise, where the thermal background of the instrument is dominant. All parameters are functions of wavelength. 

We calculated the transmission of the optical system using on-axis observations of $\kappa$ And A. We use this to find $T$. We don't expect significant overlap with stellar spectral lines in $K$-band for this stellar type, minimizing errors due to an imperfect stellar spectrum. We divide the on-axis spectra of the star by the model of the star, using a PHOENIX spectrum at 11600 K~\citep{Husser2013}, and use this to determine the transmission of planet light.

The remaining coupling terms vary slowly as a function of wavelength. As wavelength changes, differential atmospheric refraction slowly changes the apparent position on sky of the companion. Stellar speckle coupling can be approximated across $K$-band as a low-order polynomial \citep{GRAVITY2020}. These slow shifts mean that continuum subtraction using a high-pass filter can remove the effect of these terms in the data. 

Continuum subtraction also allows us to disregard the low spectral frequency changes of the instrument transmission that are introduced by changes in airmass.

We high-pass filter using a median filter on the spectrum, using a 200-pixel ($\sim$0.004 $\mu$m) box to compute and then subtract from the original.

After high-pass filtering, we are able to fit directly to our data. We further assume that, after high-pass filtering (denoted by $H$), effects shared across components are negligible, allowing us to break down our approximation further, instead filtering components individually:
\begin{equation}\label{eq:xcorr-data}
    \mathcal{H}[D_p] \approx \mathcal{H}[\alpha_p T P] + \mathcal{H}[\alpha_s T S ] + \mathcal{H}[n].
\end{equation}

We utilize this method initially described in \cite{Wang2021} to complete our further analysis, both for molecular detection and characterization of atmospheric parameters, radial velocity, and spin.

\subsection{Molecular Detection}

\begin{figure}
    \centering
    \includegraphics[width=0.45\textwidth]{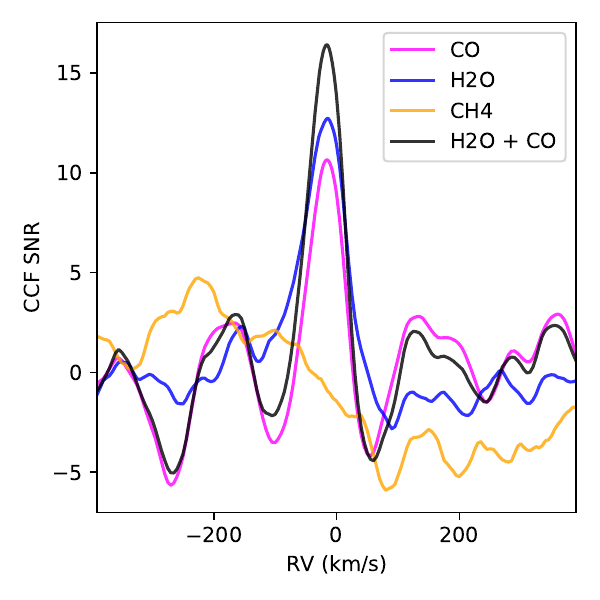}
    \caption{Cross correlations for individual molecular species. Detections of H$_2$O, CO, expected non-detection of CH$_4$. }
    \label{fig:xcorr}
\end{figure}

In order to confirm detection of the companion and identify molecules in the atmosphere of $\kappa$ And b, we used a modified cross-correlation method, estimating the maximum likelihood for both the companion and stellar flux as a function of a shift in radial velocity space, as described in~\citet{Wang2021}. For each companion RV, scaling factors for the planet and star are calculated using a linear least-square optimization technique. Our cross-correlation function (CCF) is composed of resulting values for the planet scaling parameter divided by its uncertainty at each companion RV~\citep{Ruffio2019}. From our data, we use orders 31-33 for this analysis, consistent with the locations of molecular features for  H$_2$O, CO, and CH$_4$.

We generate our planet models, based on the Sonora-Bobcat grid of temperature-pressure profiles~\citep{Marley2018, Marley2021}, allowing us to generate single molecule models for cross correlation. This is an equilibrium chemistry model without clouds, but we found the CCF signal of a combined CO+H$_2$O model to be consistent within $10\%$ of the CCF SNR, in comparison to BT-Settl models, which has clouds consistent with our expectations for this object, indicating that molecular detection is not dependent on cloud assumptions. We used model opacities from~\citet{Freedman2008, Freedman2014}, utilizing the CH$_4$ line lists from~\citet{Hargreaves2020} and H$_2$O line lists from~\citet{Barber2006}. 

Using this method, we produced single molecule models and models combining several molecules. For $\kappa$ And b, we used a model with an effective temperature of 1800~K and a surface gravity of $\log(g)$=4.5. We found these detections to be relatively insensitive to choice of effective temperature and surface gravity.

We computed CCFs using CO, H$_2$O, and CH$_4$ molecular templates for the planet data. We computed the same CCF for extracted spectra that we expected to only contain noise, both from fibers that were not on the companion but observe similar stellar flux, and from offsetting our extraction to areas of the slit without a fiber, which are dominated by the thermal background. In our CCFs, we included velocity offsets between -500 and 500 km/s from the Solar System barycenter. We normalized each CCF by dividing by the standard deviation of all noise spectral CCFs for that given molecule, giving us CCF signal-to-noise (SNR) functions.

We find a strong detection of CO and H$_2$O in the atmosphere of $\kappa$ And b, and no detection of CH$_4$. Our CCF SNR for H$_2$O alone is 12.7, CO alone is 10.6, and the combined H$_2$O and CO model is 16.4. Our strongest detection of the planet using this method was using the model containing both CO and H$_2$O. Our nondetection of CH$_4$ is consistent with the effective temperature we measure, and the width of the CCF is consistent with fast rotation. The CCF SNR functions for each molecule are plotted in Figure \ref{fig:xcorr}.

\begin{figure*}
    \centering
    \includegraphics[width=0.9\textwidth]{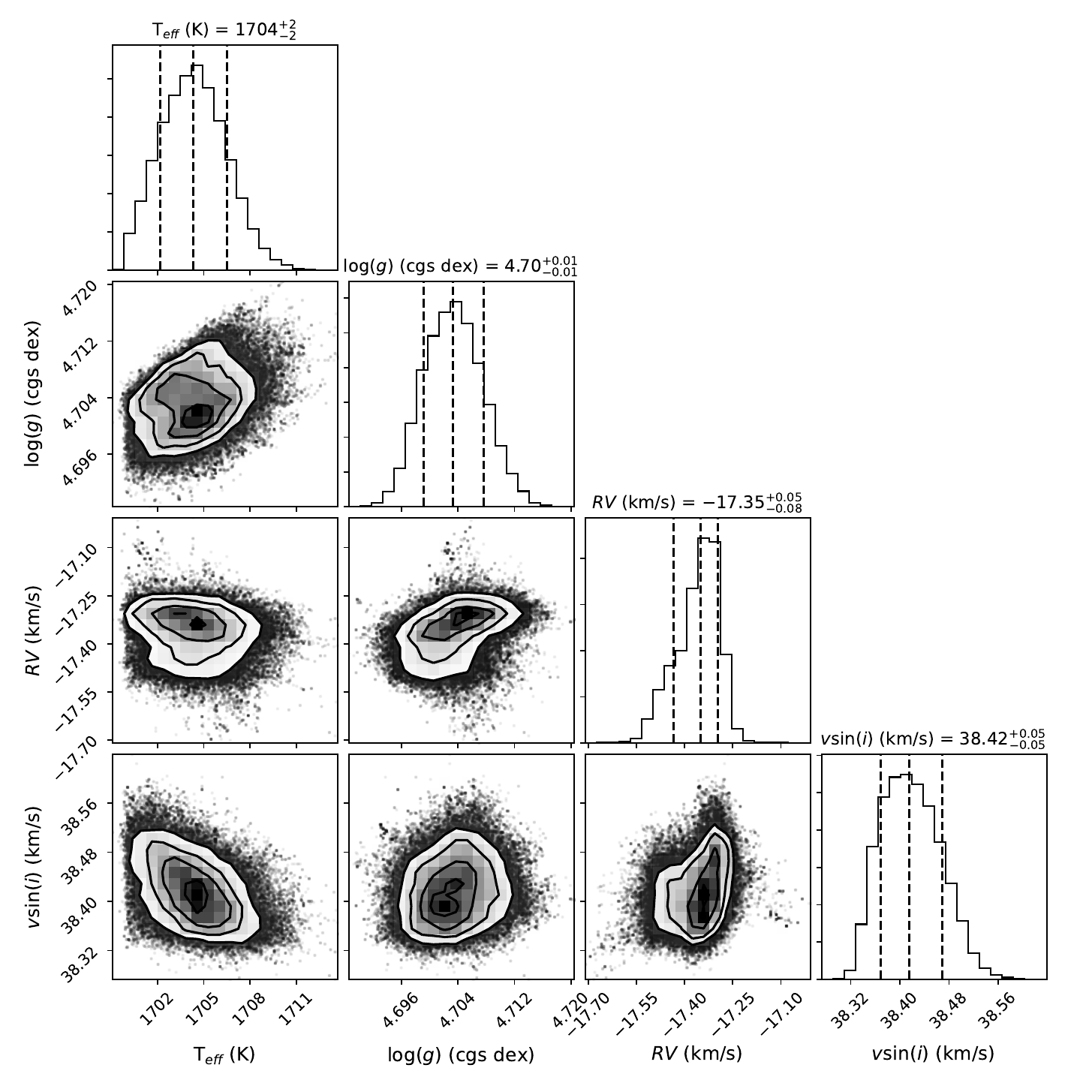}
    \caption{Corner plot corresponding to the MCMC fit to the BT-Settl model grid. Marginalized posteriors are shown on the diagonal, with covariance between all parameters in corresponding 2-d histograms. Dashed lines show 16th, 50th, and 84th percentiles, statistical error. Although this analysis results in small statistical error bars for temperature and $\log(g)$, this is not a realistic representation of our constraints on these terms. We instead adopt errors the size of the grid spacing for the rest of the paper, as our analysis strongly prefers grid points. Measurements for spin and RV are insensitive to changes of temperature or $\log(g)$ as large as the grid spacing, see Section \ref{atmosparam}.}
    \label{fig:corner}
\end{figure*}

\begin{deluxetable}{c|c|c}
\tablecaption{System Properties \label{table:sysprop}}
\tablehead{
Parameter & Measurement & Source}
\startdata
Age (Myr) & $47_{-40}^{+27}$ & \cite{Jones2016} \\
Mass ($M_\textrm{Jup}$) & $13_{-2}^{+13}$ & \cite{Currie2018} \\
Host RV (km/s) & $-12.7_{-0.8}^{+0.8}$ & \cite{Gontcharov2006} \\
Host Mass ($M_\sun$) & $2.768_{-0.1}^{+0.1}$ & \cite{Jones2016} \\
Host Spectral Type & B9IVn & \cite{Garrison1994} \\
\hline
$T_{eff}$ (K) & $1700_{-100}^{+100}$ & This work \\
$\log(g)$ (cgs dex) & $4.7_{-0.5}^{+0.5}$ & This work \\
RV (km/s) & $-17.35_{-0.09}^{+0.05}$ & This work \\
$v\sin(i)$ (km/s) & $38.42_{-0.05}^{+0.05}$ & This work \\
\enddata
\end{deluxetable}

\subsection{Companion Spin, Radial Velocity and Atmospheric Parameters}

We used the same forward modeling approach to fit for the rotational broadening  ($v\sin(i)$) of $\kappa$ And b’s spectrum using the BT-Settl model grid, as well as confirm previously constrained atmospheric and bulk parameters for the companion and measure RV.

For this analysis, we simultaneously fit the companion’s temperature, surface gravity, rotational broadening, and radial velocity relative to the Solar System barycenter, as well as the planet flux and the stellar speckles at the planet location. We fit these parameters using a Markov Chain Monte Carlo (MCMC) method built on the emcee package, which is an implementation of the affine-invariant ensemble sampler \citep{GoodmanWeare2010, ForemanMackey2013}. Our MCMC run used 100 walkers, 5000 steps, including a burn-in of 2000 steps.

For the planet spectrum component ($P$) of our forward model, we fit the BT-Settl model grid to our data, varying effective temperature ($T_\textrm{eff}$) and surface gravity ($\log(g)$ in cgs units) \citep{Allard2012}. We selected these models as BT-Settl is the only publicly available grid available at a high enough spectral resolution ($R > 35,000$) that includes clouds, present on all objects of this temperature, and confirmed present for $\kappa$ And b based on photometry and low resolution spectroscopy \citep{Stone2020, Uyama2020}. We limited by the spacing of the model grid, which offers steps of 100 K and we interpolate between, and take this as our error due to strong preference for the grid points themselves.

We fit for the planet RV relative to the Solar System barycenter, but note that the host star’s radial velocity is not well known, so relative RV measurements are difficult, but compare to existing values in the literature there as well.

The host star RV is not very well constrained because of its spectral type of B9IVn, with estimates of $-12.7\pm0.8\,\mathrm{km/s}$~\citep{Gontcharov2006} and $-11.87\pm1.53\,\mathrm{km/s}$~\citep{Becker2015}, and we use $-12.7\pm0.8\,\mathrm{km/s}$ due to the smaller uncertainty in the measurement. 

We use the \texttt{fastRotBroad} function in \texttt{PyAstronomy} \citep{pyastronomy} to fit the rotational broadening ($v\sin(i)$) of our companion spectrum.

Additionally, in preliminary analysis, we found higher than expected uncorrelated noise, than that predicted by our formal optimal extraction errors when extracting the fluxes from the traces on the detector. We believe to be caused by an underestimation of extraction errors or an unaccounted for noise term, to be explored in future analysis of instrumental noise and or data pipeline improvements. For the purpose of this work, we fit for this additional noise term in the MCMC, assuming it to be Gaussian. In this, the total error $\sigma_\textrm{tot}^2 = \sigma_\textrm{pipe}^2 + \sigma_\textrm{fit}^2$, where $\sigma_\textrm{pipe}$ is the nominal extraction error from our pipeline and $\sigma_\textrm{fit}$ is the error term we fit for. We used a separate $\sigma_\textrm{fit}$ for each spectral order, but assumed that $\sigma_\textrm{fit}$ is constant in each order. This approximation appears suitable based on analysis of fit residuals.

Our log-likelihood as described in \citet{Wang2021} is defined:
\begin{equation}
    \ln(\mathcal{L}) = -\frac{1}{2} \sum \left( \frac{(\mathcal{H}[D_p] - \mathcal{H}[M_p])^2}{\sigma_{tot}^2} + \ln{(2\pi\sigma_{tot}^2)} \right),
\end{equation}
Here, we sum over each spectral channel in our data. $M_p$ includes planet and stellar speckle parameters.

For each free parameter, we assumed a uniform prior. Effective temperature ($T_\textrm{eff}$) was assumed to be between 1000 and 2200 K, surface gravity ($\log(g)$) between 3.5 and 5.5, companion radial velocity -150 and 150 km/s, rotational broadening ($v\sin(i)$) between 0 and 80 km/s, planet flux 0 to 100 DN (data numbers, also called analog digital unit or ADU), stellar speckles flux from 0 to 500 DN for each order, and our additional noise term ($\sigma_\textrm{fit}$) from 0 to 100 DN for each order.

Results of our MCMC fits for temperature, $\log(g)$, radial velocity, and $v\sin(i)$ are shown in Figure \ref{fig:corner} and in Table~\ref{table:sysprop}. We assessed the model fit by plotting the best fitting parameters to each order of our data used in the fit, seen in Figure \ref{fig:bestfit}. On visual inspection, our forward models appear to fit the data. Residuals of the fit appear to be dominated by uncorrelated noise, which we confirm by computing the autocorrelation function (ACF) of the residuals. The ACF is well approximated by a delta function, with the wings of the ACF having an amplitude of $\leq$5\% of the peak, suggesting that an uncorrelated noise model is sufficient, likely dominated by thermal noise in the instrument. 

\section{Discussion}\label{sec:disc}

\subsection{Atmospheric Parameters}\label{atmosparam}

We detect CO and H$_2$O but not CH$_4$ in our analysis of high resolution spectra of $\kappa$ And b. This finding is consistent with previous studies of the object at lower resolution \citep{Wilcomb2020, Stone2020}. We detected each molecule independently, but found the highest correlation to a combined model.

We obtain a fit on bulk atmospheric properties from our forward model fits to the data. We find that our results are consistent with previous work. However, these findings are highly model dependent. Our effective temperature measurement of $1700_{-100}^{+100}$ K is consistent with lower resolution analysis, but prior work measures a range of temperatures as low as 1600 K and as high as 2200 K~\citep{Stone2020, Uyama2020, Wilcomb2020}. The systematic error of our fits is higher than the statistical error. Previous studies do not consistently constrain $\log(g)$, but many find in the range of 4.0 to 4.5. We find the same model dependence, with a best fit of $4.7_{-0.5}^{+0.5}$, higher than would be consistent with prior results. Discrepancies among results for $\log(g)$ are expected, as high resolution spectroscopy is not as reliable for $\log(g)$ measurements.~\citet{Wang2021} also found higher $\log(g)$ for the HR 8799 planets than in the literature from lower resolution observations.
We use the BT-Settl models, which use equilibrium chemistry, where there could be disequilibrium chemistry effects, depending on temperature. However, other publicly available models with clouds are not available at a high enough resolution, so we were not able to use these to compare.

We did not observe a strong correlation between values for radial velocity and planetary spin and our measured bulk atmospheric parameters, mitigating the propagation of issues with our atmospheric measurements into the rest of our analysis, but these inconsistencies could nevertheless bias our measurements. We fit with broad atmospheric priors to address this issue, and identify a consistent radial velocity in our single molecule fits, which are instead derived from the Sonora model grid. We also tested fixed values for Teff and $\log(g)$, finding no measurable shift in radial velocity and spin.

\subsection{Orbital Fit}

\begin{figure*}
    \centering
    \includegraphics[width=0.9\textwidth]{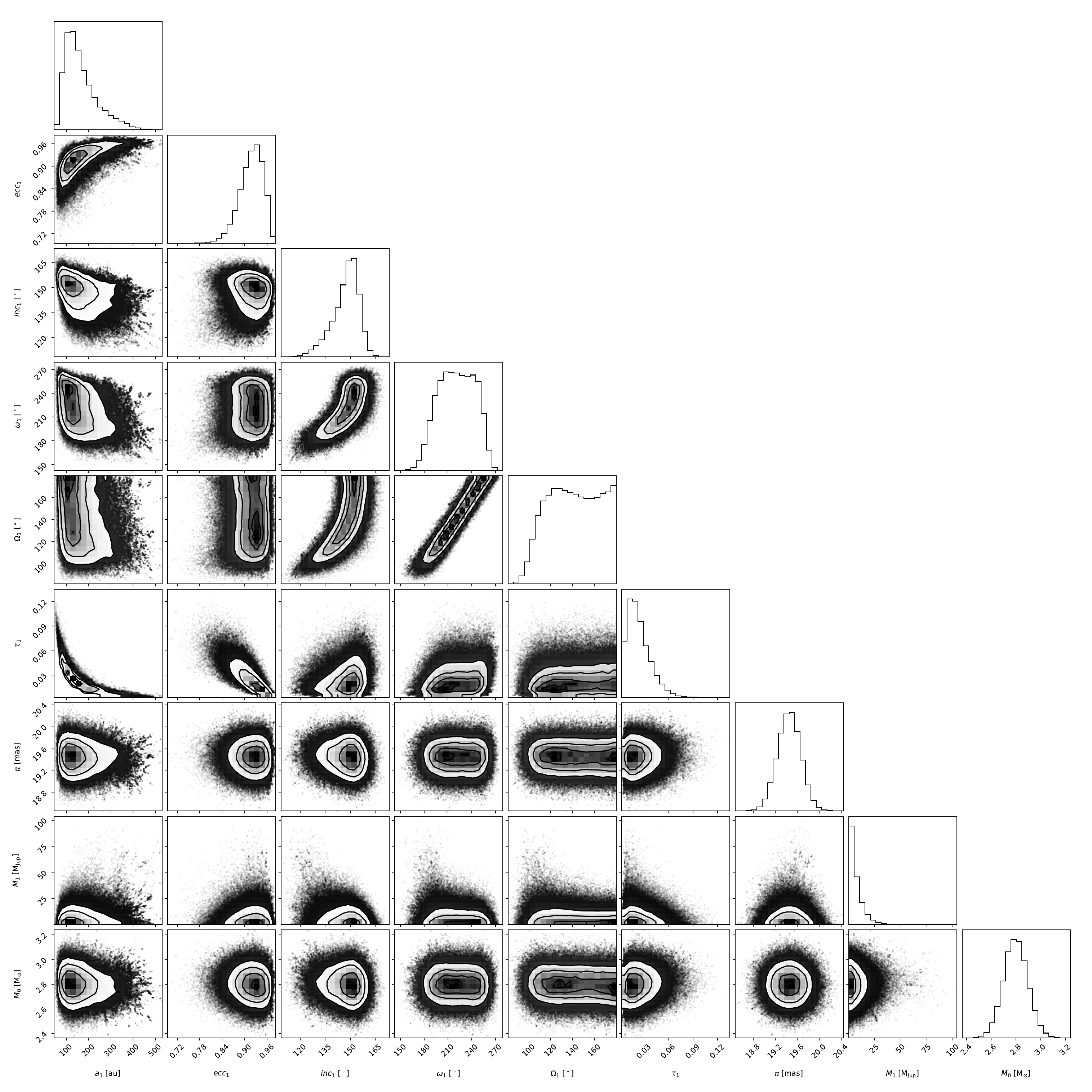}
    \caption{Posteriors of the best-fit orbital solutions for $\kappa$ And b using companion RVs, astrometry \citep{Bowler2020}, and the Hipparcos-Gaia
(EDR3) Catalog of Accelerations~\citep[HGCA;][]{Brandt2021} with the orbitize! package~\citep{Blunt2020}.}
    \label{fig:rvcorner}
\end{figure*}

\begin{figure*}
    \centering
    \includegraphics[width=0.9\textwidth]{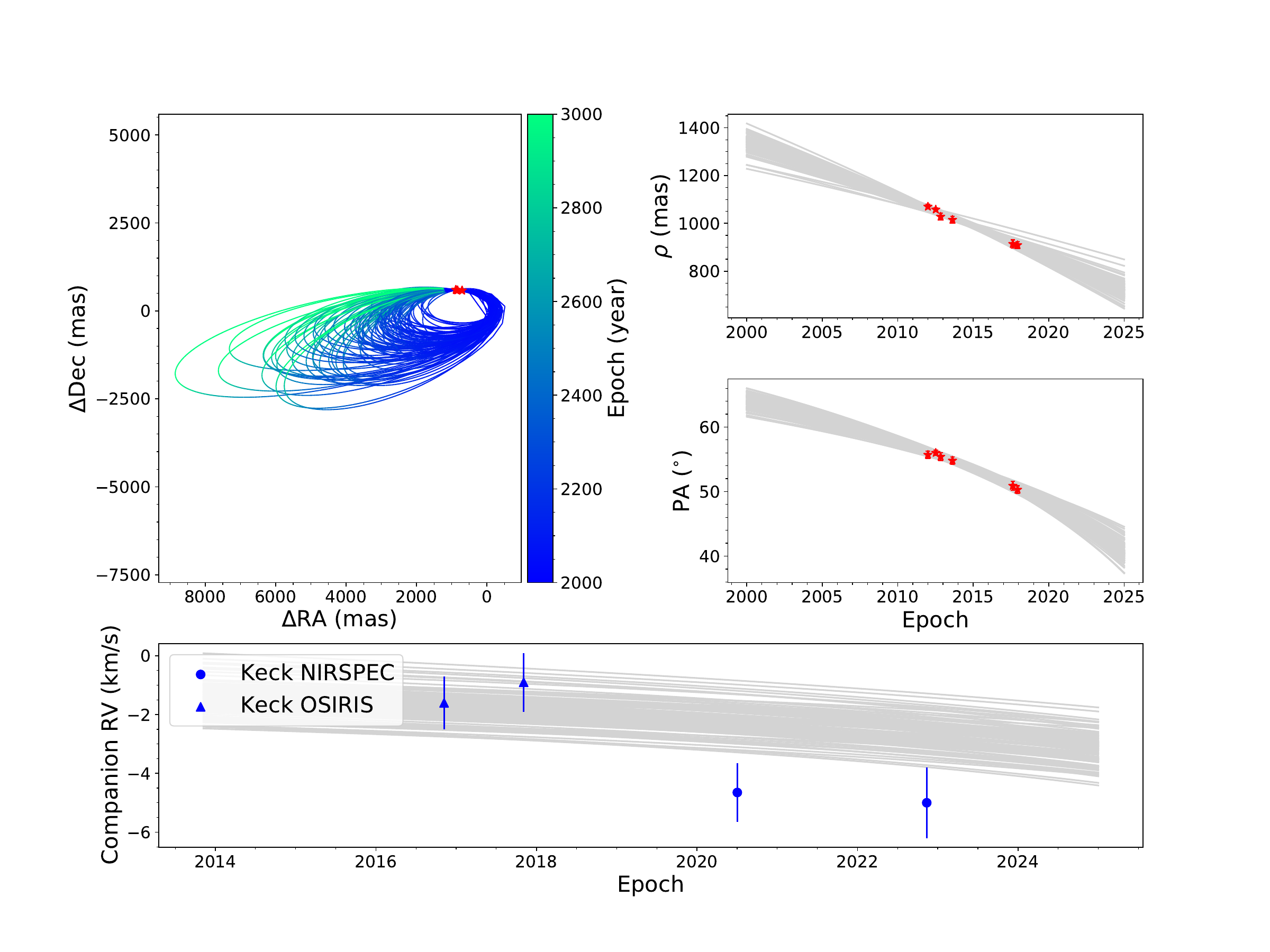}
    \caption{Best-fit 100 random draw orbits for $\kappa$ And b, using companion RVs and the Hipparcos-Gaia (EDR3) Catalog of Accelerations (HGCA; Brandt 2021) with the orbitize! package. Upper-left: $\Delta$RA and $\Delta$Dec coordinates best fits for the $\kappa$ And b system (blue/green lines) and the relative astrometry (red stars) from \citet{Bowler2020}. Upper-right; top: separation ($\rho$) in mas best fits (grey lines) with relative astrometry (red stars) from \citet{Bowler2020}; Upper-right; bottom: position angles (PA) in degrees best fits (grey lines) with the relative astrometry (red stars) from \citet{Bowler2020}. Middle: RV orbits best fits (grey lines) for the host star $\kappa$ And A. Bottom: RV orbits best fits (grey lines) for the $\kappa$ And b using KPIC measurements (blue triangles).}
    \label{fig:orbitfit}
\end{figure*}

We perform an orbit fit using existing astrometry data, compiled in \citet{Bowler2020}, with the addition of our KPIC radial velocity points. We include the radial velocity point from this study, which we measure to be $-17.35_{-0.09}^{+0.05}$~km/s, as well as another KPIC point on this object from \citet{XuanSUBMITTEDish} of $-17.7_{-0.9}^{+0.9}$~km/s, and compare to points from \citet{Wilcomb2020}. We used the orbitize! package to run an MCMC to jointly fit our radial velocity and astrometry data on the system~\citep{Blunt2020}, using the Hipparcos-Gaia
(EDR3) Catalog of Accelerations \citep[HGCA;][]{Brandt2021}. We used a parallax value of $19.4064\pm0.2104\,$ mas~\citep{GaiaEDR3} and a stellar mass value of $2.768\pm0.1\,M_\sun$~\citep{Jones2016}. We set a prior on companion mass of 0.0-0.1 $M_\sun$. Our orbitize! MCMC run used 1000 walkers, 50,000 steps, and 20 walker temperatures, to improve avoidance of local minima. Results of this run are shown in Figure \ref{fig:rvcorner} and posteriors are listed in Table~\ref{table:orbitalfit_post}. 

We find our RV measurements to fall lower than expectations for the orbit, but within our uncertainty range, seen in Figure \ref{fig:orbitfit}. Our relative RV is also affected by the significant uncertainty in the stellar radial velocity itself. Our RV addition does not significantly change or constrain the understood orbit of $\kappa$ And b. We compared to fits found in the literature, and also ran the same orbit fit excluding RV points. In comparing to \citet{Wilcomb2020}, we find that most of our overlapping parameters ($i$, $\omega$, $\Omega$) fit within the $1\sigma$ range defined there.

\begin{deluxetable}{c|c}
\tablecaption{Orbit Posteriors
\label{table:orbitalfit_post}}
\tablehead{
Parameter & Posterior}
\startdata
$a$ [AU] & $101.96_{-27.37}^{+50.36}$ \\
$e$ & $0.88_{-0.04}^{+0.03}$ \\
$i$ [deg] & $148.05_{-14.59}^{+10.94}$ \\
$\pi$ [mas] & $19.45_{-0.21}^{+0.21}$ \\
$\omega$ [deg] & $200.85_{-18.40}^{+30.81}$ \\
$\Omega$ [deg] & $115.85_{-15.46}^{+32.01}$ \\
$\tau$ & $0.04_{-0.02}^{+0.02}$ \\
$M_1$ [$M_\textrm{Jup}$] & $4.61_{-3.44}^{+7.58}$ \\
$M_0$ [$M_\sun$] & $2.79_{-0.10}^{+0.10}$ \\
\enddata
\end{deluxetable}

\subsection{Spin Analysis}

\begin{figure*}
    \centering
    \includegraphics[width=0.9\textwidth]{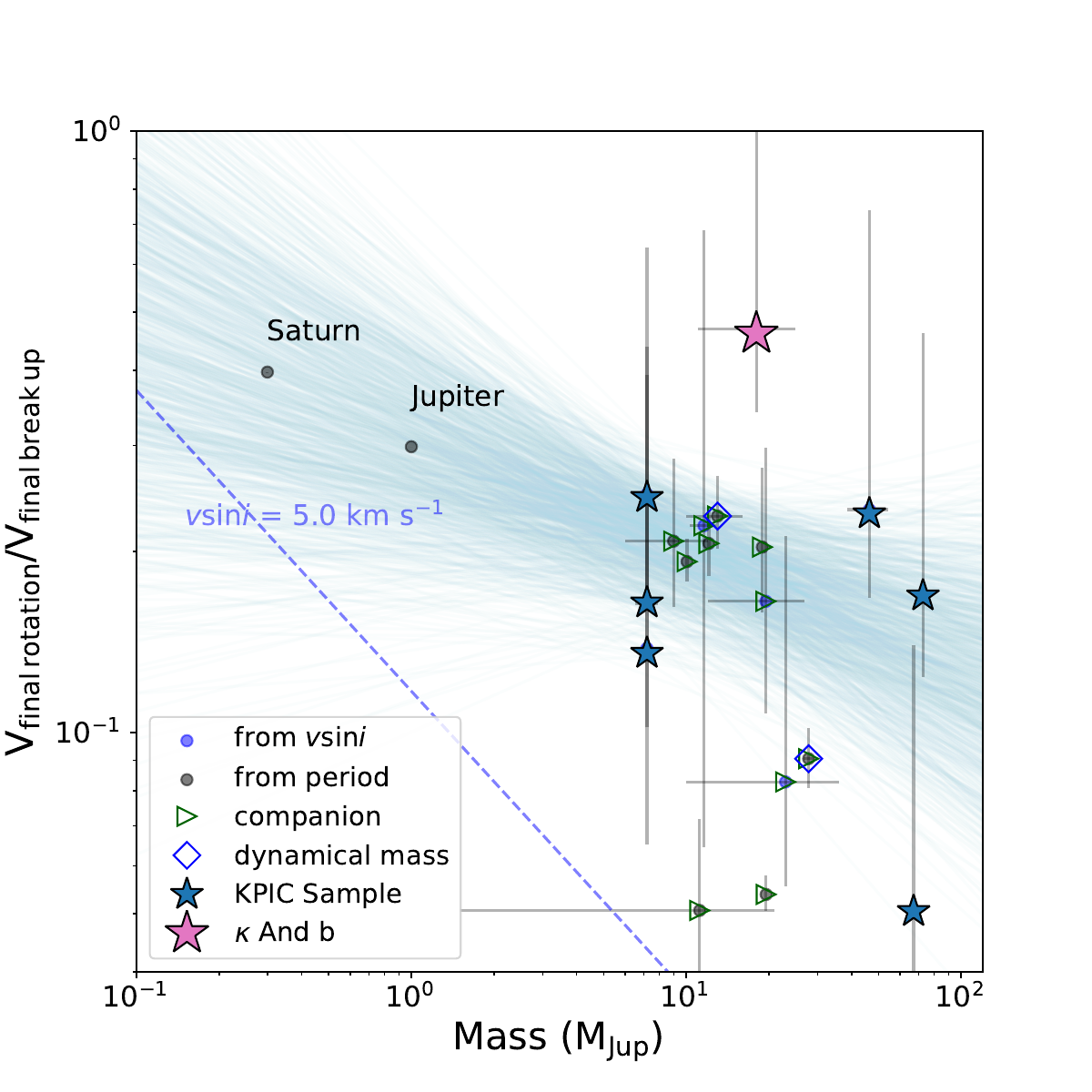}
    \caption{Spin measurements from low mass companions and solar system planets, with those from the published KPIC sample marked with a star. $\kappa$ And b, shown with a pink star, is one of the highest points in this comparison. Other known companions are marked with a green triangle, and known dynamical mass measurements with a blue diamond. Detection limits for $v\sin(i)$ ($\leq$ 5 km s-1) and photometry time series ($\leq$ 10 hr) are shown via blue and grey lines, respectively. We also include random draws from the~\citet{Wang2021} solutions for the best-fit rotational trend, in light blue.}
    \label{fig:spinanalysis}
\end{figure*}

We measured the spin of $\kappa$ And b for the first time via analysis of the $v\sin(i)$ of $38.42_{-0.05}^{+0.05}$ km/s, and compared this to a population of companions with measured spins. We compared to the population assessed in~\citet{Bryan2020}, as well as other early results from KPIC, including \citet{Wang2021, HsuNEW}. We excluded tenuous measurements. We included the rotation periods of Saturn and Jupiter in our analysis.

We use the method described in \citet{Wang2021} and \citet{HsuNEW} to convert our $v\sin(i)$ measurement to a final rotational velocity v, in order to compare it to the predicted break-up velocity of $\kappa$ And b in Figure \ref{fig:spinanalysis}. In that analysis, spins are parameterized as rotational velocity over break-up velocity. Assuming uniform inclination distributions, the geometric sine distribution is used to convert $v\sin(i)$ into a rotational velocity, and inferred age, mass, and radii are used to compute the breakup velocity. In order to minimize differences based on age in the analysis, these rotation points are evolved to 5 Gyr assuming conservation of angular momentum, as the final rotation velocity and the final breakup velocity~\citep{Wang2021,HsuNEW}. 

We see that our measurement falls within expectations for the population, though significantly on the higher end of this range, rotating at a speed much closer to the breakup velocity than most objects in the sample. The measured V$_\mathrm{final \, rotation}$/V$_\mathrm{final \, break \, up}$ for $\kappa$ And b is $0.461_{-0.127}^{+0.797}$, with a V$_\mathrm{final \, rotation}$ of $85.1_{-23.9}^{+144.8}$~km/s and a V$_\mathrm{final \, break \, up}$ of $184.9_{-22.8}^{+28.3}$~km/s. This rotation measurement is close to $50\%$ of breakup velocity, where most objects in the sample are closer to $10\%$.

We add our data point to the analysis begun in \citet{Wang2021} and continued in \citet{HsuNEW}. In this, we focused our analysis on companions that are likely to share a planet-like formation history. We did not focus on spin measurements for free-floating brown dwarfs, which may have formed through different mechanisms. However, \citet{Bryan2020} argues for a single spin regulation mechanism, independent of formation scenario. 

Additional data points allow us to improve constraints on the relationship between mass and spin in this regime, as well as investigate how spin evolves for these objects, though a strong evaluation of this relationship is not yet possible. To date, there are both few overall spin measurements, but also only a small subset of these are bound, directly imaged companions close to their host star. 

This area of parameter space is opened up significantly by KPIC, with its ability to push closer to the star. 
These initial measurements for $\kappa$ And b add important information in the higher spin velocity regime, as this object's rotation speed is one of the highest relative to breakup measured to date.

We measure $\kappa$ And b to have a $v\sin(i)$ of $38.42_{-0.05}^{+0.05}$ km/s, and put this measurement in context with the currently measured population in order to investigate a spin-mass relationship.

\section{Conclusion}\label{sec:con}

We obtained high resolution K-band spectra of $\kappa$ And b taken by KPIC. We used cross correlation methods to detect CO and H$_2$O at high significance in the planetary atmosphere, with CCF SNRs of 10.6 and 12.7 respectively and 16.4 combined, and registered a non-detection of CH$_4$.
We used a forward modeling likelihood-based framework to fit atmospheric parameters, radial velocity, and spin of the planet, while simultaneously fitting for the stellar speckles present in the data.

For effective temperature, we measure $1700_{-100}^{+100}$ K, which is limited by our model grid size. This value is consistent with some prior lower resolution findings, though these have varied widely between 1600 K and 2200 K~\citep{Stone2020, Uyama2020, Wilcomb2020}. For $\log(g)$, we measure $4.7_{-0.5}^{+0.5}$~cgs dex, higher than expected based on previous studies. We identify a radial velocity for $\kappa$ And b of $-17.35_{-0.09}^{+0.05}$ km/s, which is consistent with another KPIC data point from a separate analysis, $-17.7_{-0.9}^{+0.9}$ km/s~\citep{XuanSUBMITTEDish}, and falls slightly below our best orbit fit. We measured the spin of $\kappa$ And b for the first time at $38.42_{-0.05}^{+0.05}$ km/s, finding it to be high in comparison to the rest of the currently measured population of bound companions, closer to $50\%$ of breakup velocity, rather than the $10\%$ that is closer to standard for existing rotational measurements of bound companions.

\subsection{Future Prospects}
For $\kappa$ And b, further investigation would help particularly for studies of the planet's radial velocity. Improved information on the $\kappa$ And A would significantly improve existing and future RV points, as our error in relative RV is dominated by the error in stellar RV. More precise RV data on the planet would also improve our orbit fit.

We would also benefit from high resolution analysis of the planet at other wavelengths. Searching a wider range of wavelengths, and comparing to models with an updated line lists for CH$_4$ and other molecules, could yield detections not possible in this study, though this is not necessarily expected within the $\kappa$ And b temperature range.

Our spin trend analysis would benefit from a wider population of studied bound companions, particularly those in the planetary mass regime. This is possible with KPIC, and some studies are already underway to widen this sample~\citep{HsuNEW, XuanSUBMITTEDish}.
Future upgrades to KPIC, as well as other existing and planned instruments (current: VLT/CRIRES+~\citep{Follert2014}, VLT/HiRISE~\citep{Otten2021}, Subaru/REACH~\citep{Kotani2020}, future: Keck/HISPEC, TMT/MODHIS~\citep{Mawet2022, Konopacky2023}, ELT/ANDES~\citep{Marconi2022}) will both expand on the sample of studied bound and low-mass companions. Hardware improvements will enable greater sensitivity and higher precision, leading to better rotation, radial velocity, and atmospheric measurements.

\section*{Acknowledgements} 
Funding for KPIC has been provided by the California Institute of Technology, the Jet Propulsion Laboratory, the Heising-Simons Foundation (grants \#2015-129, \#2017-318, \#2019-1312, \#2023-4598), the Simons Foundation (through the Caltech Center for Comparative Planetary Evolution), and the NSF under grant AST-1611623. J.X. is supported by the NASA Future Investigators in NASA Earth and Space Science and Technology (FINESST) award \#80NSSC23K1434. 

Data presented in this work were obtained at the W. M. Keck Observatory, which is operated as a scientific partnership among the California Institute of Technology, the University of California and the National Aeronautics and Space Administration. The Observatory was made possible by the generous financial support of the W. M. Keck Foundation. We wish to recognize and acknowledge the very significant cultural role and reverence that the summit of Maunakea has always had within the indigenous Hawaiian community. We are most fortunate to have the opportunity to conduct observations from this mountain.

\facilities{Keck (KPIC)}

\software{\texttt{astropy} \citep{Astropy2018}, \texttt{scipy} \citep{2020SciPy-NMeth}, \texttt{PyAstronomy} \citep{pyastronomy},
\texttt{emcee} \citep{ForemanMackey2013},
\texttt{corner} \citep{corner},
\texttt{orbitize!} \citep{Blunt2020}}

\bibliography{bib_kapandb}{}
\bibliographystyle{aasjournal}

\end{CJK*}
\end{document}